\documentstyle[preprint,tighten,aps,epsfig]{revtex}

\newcommand{\kpmm}{K^+\rightarrow\pi^+\mu^+\mu^-}
\newcommand{\kmm}{K_L\rightarrow\mu^+\mu^-}
\newcommand{\beq}{\begin{equation}}
\newcommand{\eeq}{\end{equation}}
\newcommand{\bay}{\begin{eqnarray}}
\newcommand{\eay}{\end{eqnarray}}
\begin{document}

%\draft
\preprint{
\begin{tabular}{r} FTUV/99$-$17 \\ IFIC/99$-$18
\end{tabular}
}

\title{LAB observables for the muon polarization in $\kpmm$}
\author{D.\ G\'omez Dumm\thanks{now at Laboratorio Tandar,
Comisi\'on Nacional de Energ\'{\i}a At\'omica, Av.\ Libertador 8250,
(1429) Buenos Aires, Argentina} \ and \ J.\ Vidal}
\address{\hfill \\ Departament de F\'{\i}sica Te\`orica, IFIC, 
CSIC -- Universitat de Val\`encia \\
Dr.\ Moliner 50, E-46100 Burjassot (Val\`encia), Spain}
%\date{\today}

\maketitle

\begin{abstract}
We analyse the muon longitudinal polarization asymmetry $\Delta_{long}$
in the decay $\kpmm$. It is stressed that, since the muon helicities are
not Lorentz--invariant quantities, the magnitude of $\Delta_{long}$
depends in general on the reference frame. We consider the
muon helicities in the LAB system, and study the sensitivity of
the longitudinal polarization asymmetry to the flavour--mixing
parameters in the Standard Model for stopped and in--flight decaying
$K^+$. A similar analysis is carried out for the decay $\kmm$. We find
that in both cases the asymmetry is diluted when increasing the
energy of the decaying kaons.
\end{abstract}

%\pacs{PACS numbers: }

%\setcounter{page}{2}

\vspace{2cm}

\section{Introduction}

As has been pointed out some years ago by Savage and Wise~\cite{wise}, the
muon polarization in the $\kpmm$ decay can provide important information
on the structure of weak interactions and the flavour mixing. The
process is dominated by a parity--conserving contribution, arising from
the exchange of one photon. Nowadays the theoretical analysis of the
$K\rightarrow\pi\gamma^\ast$ form factor is being revisited ---including
unitarity corrections from $K\rightarrow \pi \pi \pi$ and
chiral perturbation expansion up to ${\cal O}(p^6)$ \cite{jorge}--- in
view of the recent measurement \cite{adler} of the ratio $R=\Gamma(\kpmm)/
\Gamma( K^+ \rightarrow \pi^+ e^+ e^-)$, which appears to be lower than
the prediction obtained at leading order in the chiral expansion \cite{toni}.

Parity violating observables, such as the asymmetry in the polarization of
the outgoing
$\mu^+$ and $\mu^-$, are sensitive to short-distance dynamics. In the Standard
Model (SM), the effect arises from the interference between the one--photon
amplitude and one--loop $Z$-penguin and $W$-box Feynman
diagrams~\cite{wise,bgt,buras}. It has been
shown \cite{wise} that the muon polarizations can be predicted
in terms of the well--known $K_{l3}$ semileptonic decay form factors,
and a parameter $\xi$ that carries ``clean'' information (that means,
relatively free from nonperturbative effects) on the quark masses
and mixing angles. The explicit expression of $\xi$ in terms of the quark
mixing parameters has been calculated by Buchalla and Buras~\cite{buras}
up to next--to--leading order in QCD, where the dependence on the
renormalization scale is shown to be significantly reduced.

The muon polarization asymmetry receives also potentially significant
contributions from the
interference of the one--photon amplitude with parity--violating Feynman
diagrams in which the muon pair is produced by two--photon exchange.
Though these contributions are difficult to evaluate ---they arise
from nonperturbative QCD---, a detailed analysis performed in Ref.~\cite{lu}
seems to indicate that they are smaller than the short--distance
contributions mentioned above. Here we will take this as an assumption,
focusing our attention on the effects on the muon polarization arising
from the short-distance part.

The theoretical analyses usually concentrate on the case of longitudinal
muon polarizations. This is convenient for an obvious reason, which is
the fact that the polarization direction is defined in each
case by the muon momentum, and no external axes have to be introduced.
However, the price one has to pay is that the so--defined longitudinal
polarization asymmetry $\Delta_{long}$ is not a Lorentz--invariant magnitude,
hence its value depends on the chosen reference frame. In the literature,
it is usual to define $\Delta_{long}$ in the rest frame of the $\mu^+\mu^-$
pair, and to present the theoretical results in terms
of the muon pair invariant mass, $q^2$, and $\theta$, the angle between the
three-momenta of the kaon and the $\mu^-$ in this reference frame. Then,
to compare the measurements in the LAB system with the theoretical
predictions, it is necessary not only to measure the muon polarization,
but also to reconstruct the full kinematics of each event in order to perform
the corresponding boost to the $\mu^+\mu^-$ rest frame. In addition, cuts
on the variable $\theta$, which can improve the sensitivity to the 
short--distance parameter $\xi$ mentioned above \cite{lu}, do not translate,
in general, into cuts on the pion and muon directions in the LAB.
All these facts
produce additional sources of uncertainties in the analysis. The aim of
this work is to point out these difficulties, and propose the longitudinal
polarization asymmetry defined in the LAB system,
$\Delta_{long}^{({\rm LAB})}$, as the best observable
to be contrasted with experiment. We analyse here the kinematics for the
process in the LAB frame, and calculate the expected sensitivity of
$\Delta_{long}^{({\rm LAB})}$ to the parameter $\xi$, for both stopped and
in--flight decaying kaons. For a fixed energy of the $K^+$, we show that the
sensitivity of the observable can be improved by a convenient cut on the LAB
muon energy. In addition, we perform a similar analysis for the decay $\kmm$.
The study of the muon polarization is also important in this case, since it
can provide a new signal of CP violation~\cite{cpkmm}. From our analysis,
it arises that the asymmetry $\Delta_{long}^{({\rm LAB})}$ is partially
diluted when the decaying kaons are in flight.

The paper is organized as follows: in Section II we study the 
sensitivity of $\Delta_{long}^{({\rm LAB})}$ to the SM parameters
for the process $K^+ \rightarrow \pi^+\mu^+\mu^-$, and calculate
the dependence of the observable with the $K^+$ energy.
Then, in Section III, we perform a similar analysis for the
process $K_L\rightarrow \mu^+ \mu^-$, in which the kinematics is simpler.
In section IV we present our conclusions. Details on the LAB frame
kinematics and phase space integrations are given in the Appendix.

\section{Muon polarization and kinematics for $\kpmm$}

As stated, the decay rate for $\kpmm$ is dominated by the one-photon
exchange contribution, which is parity-conserving. The corresponding
amplitude can be parametrized as
\beq
{\cal M}^{(PC)}=\frac{\alpha G_F\sin\theta_C}{\sqrt{2}} f(q^2)
(p_K+p_\pi)^\mu \bar u(p_-,s_-)\gamma_\mu v(p_+,s_+)\,,
\label{pc}
\eeq
where $p_K$, $p_\pi$ and $p_\pm$ are the four--momenta of the kaon, pion
and $\mu^\pm$ respectively, and $q^2=(p_++p_-)^2$ stands for the
squared $\mu^+\mu^-$ invariant mass. We consider the general case of
polarized muons, being $s_\pm$ the corresponding polarization vectors.

In the Standard Model, in addition to the dominant term (\ref{pc}),
the decay amplitude contains a parity--violating piece. This
can be written in general as
\beq
{\cal M}^{(PV)}=\frac{\alpha G_F\sin\theta_C}{\sqrt{2}} \left[B
(p_K+p_\pi)^\mu +C (p_K-p_\pi)^\mu\right]
\bar u(p_-,s_-)\gamma_\mu\gamma_5 v(p_+,s_+)\,,
\label{pv}
\eeq
where the parameters $B$ and $C$ get contributions from both
short-- and long--distance physics. The short--distance contributions
arise mainly from $Z$-penguin and $W$-box Feynman diagrams and
carry clean information on the flavour structure of the SM. Hence,
the experimental determination of $B$ and $C$ would be very interesting
from the theoretical point of view,
provided that the long--distance effects are under control.
Since the total decay amplitude is dominated by the parity--conserving
piece, to get this information one is lead to search for a
parity--violating observable. The
muon polarizations are immediate candidates in this sense.

It can be seen that, from the experimental point of view, the measurement
of the $\mu^+$ polarization is strongly favoured in comparison with that
of the $\mu^-$. The reason is that the $\mu^-$ give rise to the formation
of muonic atoms when they are stopped in materials, and this makes it
difficult to measure the polarization~\cite{kuno}. We will concentrate
then in the polarization
of the outgoing $\mu^+$, summing over the final $\mu^-$ states.
In an arbitrary reference frame, the decay rate for polarized $\mu^+$ is
given by
\beq
\Gamma(s_+) = \frac{1}{2E_K} \int d\Phi \sum_{s_-}|{\cal M} (s_+)|^2\;,
\label{gdef}
\eeq
where $d\Phi$ is the Lorentz--invariant differential phase space,
\beq
d\Phi=(2\pi)^4\delta^{(4)}(p_K-p_\pi-p_+-p_-)\prod_{a=\pi,+,-}
\frac{d^3p_a}{(2\pi)^3 2E_a}\;.
\eeq
The polarization asymmetry of the outgoing $\mu^+$, in the direction
given by $s_+$, is defined now as
\beq
\Delta\equiv \frac{\Gamma(s_+)-\Gamma(-s_+)}{\Gamma(s_+)+\Gamma(-s_+)}\;.
\label{obs}
\eeq
As long as $s_+$ transforms as a four--vector, this quantity is clearly
Lorentz--invariant. However, one has to take some care when referring to
the {\em longitudinal} or {\em transverse} muon polarizations. A
polarization vector that is parallel to the muon momentum in
a particular frame, acquires in general a transverse component when one
moves to a boosted system. Hence, in principle, the value of the
longitudinal polarization asymmetry $\Delta_{long}$
obtained in an arbitrary reference frame may change, or even vanish,
when the longitudinal polarization is measured in
the LAB frame. The same can be applied to transverse polarizations,
which also require the introduction of an additional reference plane
({\em e.g.}\ the plane of the decay, in the rest frame of the $K^+$).

It is usual (see for instance Refs.\ \cite{wise,buras,lu}) to define
$\Delta_{long}$ in the rest frame of the $\mu^+\mu^-$ pair. As commented
in the introduction, this represents
a problem from the experimental point of view, since the
theoretical prediction obtained for the asymmetry cannot be contrasted
by measuring only the longitudinal muon polarizations in the LAB frame. One
should instead fully reconstruct each observed event
(not just look at the final $\mu^+$), in order to boost all decay products
to the particular frame proposed, and then perform the comparison with
the theoretical value. Or, alternatively, one can boost the
polarization vector $s_+$, defined to be parallel to the $\mu^+$ momentum
in the rest frame of the $\mu^+\mu^-$ pair, to the LAB frame.
Then, to compare with the theoretical prediction, one would have to measure
the LAB $\mu^+$ polarization along a different axis for each individual
event, this axis being determined by the boost.
Once again the analysis turns out to be quite involved.

Our proposal is simple: it just consists in considering an observable
equivalent to (\ref{obs}), but defined directly in the LAB system.
The advantage is that, once the process has been identified,
the theoretical prediction
can be contrasted just by analysing the final $\mu^+$ polarization,
without taking care about the energy and angular distribution of the
remaining $\mu^-$ and $\pi^+$. In particular, we can take $s_+$ to be
longitudinal in the LAB system, and calculate the value for
$\Delta_{long}^{({\rm LAB})}$. As we show below, the result is in general
different from that obtained in the $\mu^+\mu^-$ rest frame, and
depends on the energy of the decaying $K^+$.

The detailed calculation of the decay rate for polarized $\mu^+$ in the
LAB reference frame is presented in the Appendix. We end up with the
following expression:
\beq
\Gamma(s_+) = \frac{(\alpha G_F\sin\theta_C)^2}{16\pi^2 E_K}
\int_{E_{min}}^{E_{max}} dE_+\int_{h(E_+)}^1 d(\cos\theta)
|\vec p_+| \left[g_0(z)+ (s_+\cdot p_K) g_1(z) \right]\,,
\label{gamma}
\eeq
where $z=(p_K\cdot p_+)=E_K E_+ - |\vec p_K||\vec p_+| \cos\theta$, and
the integration limits $E_{min}$, $E_{max}$ and $h(E_+)$ are functions
of the Lorentz factor $\gamma$ characterising the boost from the $K^+$
rest frame to the LAB system ($\gamma=E_K/m_K$). The functions
$g_{0,1}(z)$, given in the Appendix, carry the information on the form
factors $f$, $B$ and $C$ introduced in Eqs.\ (\ref{pc}) and (\ref{pv}).

We concentrate on the longitudinal $\mu^+$ polarization,
$\Delta_{long}^{({\rm LAB})}$, which can be trivially
obtained from (\ref{obs}) and (\ref{gamma}) by taking
\beq
%s_+^0=\frac{|\vec p_+|}{m_\mu}\,,\quad\quad\quad
%\vec s_+=\frac{E_+}{m_\mu}\, \frac{\vec p_+}{|\vec p_+|}\,,
s_+^\alpha=\frac{1}{m_\mu}\left(|\vec p_+|\;,\frac{E_+}{|\vec p_+|}
\,\vec p_+\right)\;,
\label{polvec}
\eeq
with $E_+$, $\vec p_+$ in the LAB frame. Now, in order to determine the
sensitivity of this observable to the parameters of interest,
we need some theoretical input for the form factors $f$, $B$ and $C$.
In the case of $f(q^2)$, which corresponds to the effective vertex
$K\pi\gamma^\ast$, one can use the experimental information
from the decay $K^+\rightarrow\pi^+e^+e^-$. It is seen \cite{kpee}
that the absolute value of this form factor can be approximated by
\beq
|f(q^2)|=|f(0)|\left(1+\lambda\frac{q^2}{m_\pi^2}\right)\;,
\eeq
with $|f(0)|=0.294$ and $\lambda=0.105$. On the other hand, from
existing analyses within Chiral Perturbation Theory~\cite{toni,lu},
one expects the imaginary part of $f(q^2)$ to be negligibly small
compared with the real part.

In the case of the parity--violating amplitude, the situation is more
complicated due to the interference between
short-- and long--distance contributions. As stated above, the long--distance
effects arise from nonperturbative QCD and are very difficult to
estimate~\cite{lu}. To get definite numerical results, we will concentrate
here only on the effect produced by the short--distance part
(in fact, the estimates in Ref.~\cite{lu} indicate that this should be the
dominant one), in which the $q^2$ dependence of $B$ and $C$ can be
obtained from semileptonic kaon decays. We have thus
\beq
B=f_+(q^2)\,\xi\;,\quad\quad\quad
C=\frac{1}{2} f_-(q^2)\,\xi\;,
\eeq
where $f_+(q^2)$ and $f_-(q^2)$ are the well--known form factors for
$K_{l3}$ decays. We will use here a standard parametrization~\cite{kl3},
taking
\beq
f_\pm (q^2)=f_\pm(0) \left(1+\lambda_\pm\frac{q^2}{m_\pi^2}\right)\;,
\eeq
with $f_+(0)=0.99$, $\lambda_+=0.03$, $f_-(0)=-0.33$ and $\lambda_-=0$.
The novel information is contained in $\xi$, which can be calculated in
the SM in terms of the quark masses and mixing
angles. One has~\cite{wise,lu}
\begin{equation}
\xi\simeq -\widetilde{\xi}_c+
\left[\frac{V^\ast_{ts}V_{td}}{V^\ast_{us}V_{ud}}\right]
\widetilde{\xi}_t \;,
\end{equation}
where $V$ stands for the Cabibbo--Kobayashi--Maskawa matrix, and
$\widetilde{\xi}_c$ and $\widetilde{\xi}_t$ arise from the contributions
of $Z$-penguins and $W$-boxes. QCD corrections introduce some dependence
on the renormalization scale, though this can be reduced with the
inclusion of next--to--leading order contributions~\cite{buras}.

Here we just keep $\xi$ as a parameter, and refer the reader to
Ref.~\cite{buras} for the detailed analysis of its explicit dependence
on the quark masses and mixing angles. Since both $B$ and $C$ are linear
in $\xi$, the muon longitudinal polarization asymmetry can be written as
\beq
\Delta_{long} = \pm{\rm Re}\xi\, R\;,
\eeq
where the $\pm$ signs correspond to $|f(0)|=\pm f(0)$ respectively.
To see the sensitivity of $\Delta_{long}$ to $\xi$, we concentrate
on the value of the ``kinematic'' factor $R$, which can be computed
numerically using the above inputs for the form factors.

We recall that $\Delta_{long}$, and thus $R$, depend in general on the
reference frame in which the polarization vectors are defined to be
parallel to the $\mu^+$ momenta. In the LAB system, $R$ can be
calculated by means of Eqs.\ (\ref{gamma}) and (\ref{polvec}) in terms of
the energy of the decaying $K^+$. The resulting curve is shown in
Fig.\ 1. It can be seen that the asymmetry is maximized when the
kaons are at rest, with $R\simeq -2.9$, while the effect turns out
to be diluted for in--flight $K^+$. For a dilation factor
$\gamma\rightarrow\infty$ we end up with $R\simeq -1.6$.

In the case of high--energy kaons the sensitivity can be improved by
performing a convenient cut in the $\mu^+$ energy. We have analysed the
situation for a dilation factor $\gamma=12$, this means, a kaon
energy of about 6 GeV. This is the energy of the $K^+$ beam in the
experiment E865 at the BNL AGS, used to study $K^+$ decays involving three
charged final particles, and suggested as one of the best
candidates to perform the measurement of the $\mu^+$ polarization in
$\kpmm$ \cite{report}. The dependence of the asymmetry with the
chosen range of the outgoing $\mu^+$ energy for $\gamma=12$
can be seen from Fig.\ 2, where we plot the differential rates
\[
\frac{d\Gamma(s_+)}{dE_+}+\frac{d\Gamma(-s_+)}{dE_+}
\;,\quad\quad
\frac{1}{{\rm Re}\xi}\left(
\frac{d\Gamma(s_+)}{dE_+}-\frac{d\Gamma(-s_+)}{dE_+}\right)
\]
---the latter, up to a global sign--- in terms of the $\mu^+$ energy
$E_+$. As it is shown in the figure, there is a change of sign in the
$\mu^+$ polarization for $E_+\sim 1$ GeV. This leads
to a reduction in the value of $|R|$ when integrating over the whole
range of $\mu^+$ energies. By taking a lower cut at $E_+=1$ GeV,
the asymmetry increases from $R\simeq -1.6$ to $-2.1$, while the number
of events gets reduced only by a factor 0.82.

Finally we notice that, by working in the $\mu^+\mu^-$ rest frame, one
obtains $|R|=2.3$~\cite{lu}. Thus the best sensitivity for
$\Delta_{long}$, with no phase--space cuts, would be obtained from the
decay of stopped kaons.

\section{Muon polarization and kinematics for $\kmm$}

The above discussion about the fermion polarizations and the dependence
on the reference frame can be also applied to the decay $\kmm$.
In this case, the longitudinal polarizations of the outgoing muons
have also a considerable theoretical interest, since
the measurement of nonzero polarizations would represent a new
signal of CP violation~\cite{cpkmm}. It is clear that, being
$\kmm$ a two--body decay, the kinematics is now much simpler than in
the $\kpmm$ case.

Using a similar notation as in the previous section, the decay
amplitude for $\kmm$ can be written in terms of two parameters
$A$ and $B$,
\beq
{\cal M}=\bar u(p_-,s_-)\,(i\, B+ A \gamma_5)\, v(p_+,s_+)\,.
\label{ampkmm}
\eeq
To study this process, it is usual to work in the kaon rest frame,
where the analysis is simpler. One can define the longitudinal
$\mu^+$ polarization asymmetry $\Delta_{long}^{({\rm rest})}$ by using an
expression similar to (\ref{obs}), and taking the polarization
vectors $\vec s_+$ to be parallel to the $\mu^+$ three--momenta in the kaon
rest frame. As we have discussed below, $\Delta_{long}^{({\rm rest})}$ will
in general be different from $\Delta_{long}^{({\rm LAB})}$, defined
by taking $\vec s_+$ parallel to $\vec p_+$ in the LAB system, if the
decaying $K_L$ are in flight.

Let us analyse the dependence of $\Delta_{long}^{({\rm LAB})}$ with the
energy of the decaying $K_L$, $E_K=\gamma m_K$. As before, we sum over the
final $\mu^-$ polarizations, obtaining
\beq
\sum_{s_-}
|{\cal M}|^2 = m_K^2 \left(|A|^2+\beta_0^2 |B|^2\right)
+ 4 m_\mu\, {\rm Im}(BA^\ast) (s_+\cdot p_K)\,,
\label{msqkmm}
\eeq
where $\beta_0=(1-4m_\mu^2/m_K^2)^{1/2}$, and the factor Im$(BA^\ast)$
carries the CP violation effects. For this process the integration over
the phase space is straightforward,
and we can work directly in the LAB system. The decay rate for polarized
$\mu^+$ is found to be
\beq
\Gamma(s_+)=\frac{m_K^2}{16\pi E_K |\vec p_K|}
\int^{E_{max}}_{E_{min}} d E_+\,
\left[|A|^2+\beta_0^2 |B|^2+\frac{4 m_\mu}{m_K^2} (s_+\cdot p_K)
\,{\rm Im}(BA^\ast)\right]\,,
\label{gamkmm}
\eeq
where the limits of integration are
\beq
E_{min}= \frac{E_K-\beta_0|\vec p_K|}{2}\,,\quad\quad\quad
E_{max}= \frac{E_K+\beta_0|\vec p_K|}{2}\;.
\eeq
In the case of longitudinal polarization vectors, the scalar product
in (\ref{gamkmm}) is given by
\beq
(s_+\cdot p_K)=\frac{1}{m_\mu|\vec p_+|}\left(\frac{E_+ m_K^2}{2}
-E_K m_\mu^2\right)\;,
\label{sporp}
\eeq
and we obtain for the total $\mu^+$ polarization
\beq
\Delta_{long}^{({\rm LAB})} =\frac{{\rm Im}(BA^\ast)}{|A|^2+\beta_0^2 |B|^2}
\,\left[\frac{2\beta'}{\beta\beta_0}-\frac{(1-\beta_0^2)}{\beta\beta_0}
\log\left(\frac{1+\beta'}{1-\beta'}\right)\right]\;,
\label{labkmm}
\eeq
where
\beq
\beta=\sqrt{1-\gamma^{-2}}\;,\quad\quad
\beta'=\left\{
\begin{array}{clc}
\beta & , \hspace{.7cm} & \gamma < \frac{m_K}{2m_\mu} \\
\rule{0cm}{0.8cm} \beta_0\;\; & , & \gamma \geq \frac{m_K}{2m_\mu}
\end{array}
\right.\;.
\eeq
Notice that the dependence of the observable in Eq.\ (\ref{labkmm}) with the
$K_L$ energy is contained into the factor in square brackets, hence it
does not depend on the dynamics. As in the case of $\kpmm$, it is seen
that the asymmetry is reduced when
the kaons are more energetic, though the effect is rather small.
In the limit $\gamma\rightarrow\infty$, the longitudinal polarization
is reduced by a factor
\beq
r\equiv\frac{\Delta_{long}^{(\gamma\rightarrow\infty)}}
{\Delta_{long}^{({\rm rest})}} =
\frac{1}{\beta_0}-\frac{(1-\beta_0^2)}{2\beta_0^2}\,
\log\left(\frac{1+\beta_0}{1-\beta_0}\right)
\simeq  0.77 \;.
\eeq
Still this ratio can be slightly increased by taking into account the
energy distribution of the muons in the LAB system. Since
the CP--conserving terms in Eq.\ (\ref{msqkmm}) are independent of the
kinematic variables, the dependence of the $\mu^+$ polarization
with $E_+$ is basically given by the scalar product (\ref{sporp}).
For $E_K > m_K^2/(2 m_\mu)$, it is seen that the polarization changes
sign at $E_+ = E_0 \equiv 2E_K m_\mu^2/m_K^2$, thus the sensitivity can
be improved by making a lower cut on $E_+$. Taking
{\em e.g.}\ $E_+\geq 2E_0$, one gets $r=0.89$, while the number of events
is reduced by about 15\%.

\section{Conclusions}

We have analysed the decays $\kpmm$ and $\kmm$, focusing our attention on
the longitudinal polarization of the outgoing $\mu^+$. For these processes,
the asymmetry in the production of muons with opposite helicities
has a significant theoretical interest in connection with the flavour
mixing and the structure of the Standard Model.

The longitudinal polarization asymmetry $\Delta_{long}$ depends in
general on the chosen reference frame, since the helicity of a massive
particle can change after a Lorentz transformation. Here we have considered
$\Delta_{long}^{({\rm LAB})}$, that means, the longitudinal
polarization asymmetry defined in the laboratory system. For the decay
$\kpmm$, the advantage of
choosing this frame is that the theoretical predictions can be contrasted
with experiment just by measuring the polarization of the outgoing
$\mu^+$, summing over all energies and angular distributions of
the remaining $\mu^-$ and $\pi^+$.

For both processes, we analyse the dependence of
$\Delta_{long}^{({\rm LAB})}$ with the energy of the decaying kaons,
showing that the asymmetry is partially diluted when the kaons
are in flight. In the case of the decay $\kpmm$, this is illustrated
by the curve in Fig.\ 1 (we have neglected here long--distance
contributions arising from two--photon exchange). We have considered
in particular the case of in--flight kaons with energy of 6 GeV.
For this energy, it is shown that there is a change of sign in the
$\mu^+$ polarization for $\mu^+$ energies of about 1 GeV, thus a lower
energy cut at this point allows to improve the asymmetry.
On the other hand, in the case of the decay $\kmm$ it is shown that
the dilution is purely kinematic, {\em i.e.} it does not depend at
all on the dynamics of the process. In the limit of large $K_L$ energies,
the asymmetry $\Delta_{long}^{({\rm LAB})}$ is found to be reduced by
about 23\% with respect to the value obtained when the decaying kaons
are at rest.

\acknowledgements

We thank J.\ Bernab\'eu for useful discussions and A.\ Pich and
J.\ Portol\'es for the critical reading of the manuscript.
D. G. D.\ has been supported by a grant from the Commission
of the European Communities, under the TMR programme (Contract
N$^\circ$ ERBFMBICT961548). This work has been funded by
CICYT (Spain) under the Grant AEN-96-1718 and by DGEUI
(Generalitat Valenciana, Spain) under the Grant GV98-01-80.

\appendix

\section*{}

We calculate here the decay rate $\Gamma(s_+)$ for the process $\kpmm$
in the LAB reference frame. From Eqs.\ (\ref{pc}), (\ref{pv}) and
(\ref{gdef}), we have
\beq
\Gamma(s_+) = \frac{(\alpha G_F\sin\theta_C)^2}{2E_K}
\int\, d\Phi \left[F_0 + (s_+\cdot T)\right]\;,
\label{sqamp}
\eeq
where
\bay
F_0 & = & |f(q^2)|^2 [ 2 (2z-q^2) (m_K^2-2z) -4z m_\pi^2] \,,
\nonumber\\
T^\mu & = & \mbox{Re}(f(q^2)B^\ast) \left[
(2m_K^2-2m_\pi^2+q^2-4z)p_K^\mu+2(z-m_K^2)p_-^\mu\right]\nonumber\\
& & + \mbox{Re}(f(q^2)C^\ast)(q^2 p_K^\mu-2z p_-^\mu)\,,
\label{terminos}
\eay
with $z \equiv (p_K\cdot p_+)$.

Let us first perform the integration over the $\mu^-$ and $\pi^+$ phase
space variables. To do this, we write the differential phase space as
\[
d\Phi=\frac{d^3p_+}{(2\pi)^3 2E_+}d\Phi'\;,
\]
with
\[
d\Phi'=(2\pi)^4\delta^{(4)}(p_K-p_\pi-p_+-p_-)\,
\frac{d^3p_-}{(2\pi)^3 2E_-} \,
\frac{d^3p_\pi}{(2\pi)^3 2E_\pi}\;.
\]
Notice that $F_0$ is a function of the invariants $q^2$ and $z$. Then the
integral over the $\mu^-$ and $\pi^+$ momenta must be a function
of $z$ only,
\beq
g_0(z)=\int F_0(z,q^2) d\Phi'\;.
\label{fcero}
\eeq
In the same way, for the second term in the integrand of
(\ref{sqamp}) we can write
\beq
\int T^\mu d\Phi' = g_1(z)\, p_K^\mu+g_2(z)\, p_+^\mu \;.
\label{tmu}
\eeq
Since $s_+$ is by definition orthogonal to $p_+$, the function $g_2(z)$
does not contribute to the decay rate (\ref{sqamp}) and we only need
to compute $g_1(z)$. The latter can be written as
\beq
g_1(z) = \int F_1(z,q^2) d\Phi' \;,
\label{g1}
\eeq
where the integrand is given by
\beq
F_1(z,q^2) = \frac{\left[z\,(p_+ \cdot T)
-m_\mu^2 (p_K\cdot T)\right]}{z^2-m_\mu^2m_K^2}\;.
\eeq
To perform the integrals in (\ref{fcero}) and (\ref{g1}) explicitly, one
can choose a convenient reference frame. Let us consider the system
in which the kaon and the $\mu^+$ three--momenta have
equal magnitude and direction:
\beq
\vec p_K-\vec p_+ =  \vec p_-+\vec p_\pi = 0 \;.
\eeq
Denoting by $y$ the cosine of the angle between the $K^+$ and $\pi^+$
directions in this frame, the functions $g_{0,1}(z)$ can be obtained
from
\beq
g_i(z) = \int F_i(z,q^2) d\Phi'=
\frac{1}{16\pi} \frac{\left[(m_K^2-m_\pi^2-2 z)^2
-4 m_\mu^2m_\pi^2\right]^{1/2}}{m_K^2+m_\mu^2-2 z} \int_{-1}^1
F_i(z,q^2) dy\;,
\label{intf}
\eeq
where the $\mu^+\mu^-$ invariant
mass $q^2$ is given in terms of $z$ and $y$ by
\bay
q^2 & = & m_K^2+m_\pi^2-\frac{(m_K^2-z)(m_K^2+m_\pi^2-2z)}{m_K^2+m_\mu^2-2z}
\nonumber \\
& & + \frac{(z^2-m_K^2m_\mu^2)^{1/2} [(m_K^2-m_\pi^2-2z)^2-
4m_\pi^2m_\mu^2]^{1/2}}{m_K^2+m_\mu^2-2z}\, y\;.
\eay
The advantage of choosing this particular reference frame is that the
integration limits for $y$ do not depend on the $K^+$ or $\mu^+$ momenta.
These only enter through the Lorentz invariant product $z=(p_K\cdot p_+)$,
which does not depend on $y$.

Being $g_0(z)$ and $g_1(z)$ Lorentz--invariant functions, we can move now
easily to the LAB reference frame. The total $\kpmm$ decay rate
for polarized $\mu^+$ will be given by
\bay
\Gamma(s_+) & = & \frac{1}{2E_K}(\alpha G_F\sin\theta_C)^2
\int\frac{d^3p_+}{(2\pi)^3 2E_+}\left[g_0(z)+
 (s_+\cdot p_K) g_1(z) \right] \nonumber \\
& = & \frac{(\alpha G_F\sin\theta_C)^2}{16\pi^2 E_K}
\int_{E_{min}}^{E_{max}} dE_+\int_{h(E_+)}^1 d(\cos\theta)
|\vec p_+| \left[g_0(z)+ (s_+\cdot p_K) g_1(z) \right]
\label{gamapp}
\eay
where $\theta$ stands for the angle between the $\mu^+$ and the kaon in
the LAB system, and $z$ is given by
\beq
z=E_K E_+ - |\vec p_K||\vec p_+| \cos\theta\,.
\eeq
The limits of integration in (\ref{gamapp}) are found to be
\bay
\rule{0cm}{1.3cm} E_{min} & = & \left\{
\begin{array}{clc}
m_\mu & , \hspace{.7cm} & \gamma\leq \frac{E_0}{m_\mu} \\
\rule{0cm}{0.7cm}
\gamma E_0 -\gamma\beta |\vec p_0| & , & \gamma > \frac{E_0}{m_\mu}
\end{array}
\right. \nonumber \\
\rule{0cm}{0.9cm} E_{max} & = & \gamma E_0 + \gamma\beta |\vec p_0|
\nonumber \\
\rule{0cm}{1.5cm} h(E_+) & = & \left\{
\begin{array}{clr}
-1 & , \hspace{.7cm} &
m_\mu\leq E_+\leq \gamma E_0 -\gamma\beta |\vec p_0| \\
\rule{0cm}{0.8cm} \frac{\gamma E_+-E_0}{\gamma\beta\sqrt{E_+^2-m_\mu^2}}
& , & \gamma E_0 -\gamma\beta |\vec p_0| < E_+
\leq \gamma E_0 +\gamma\beta |\vec p_0|
\end{array}
\right.
\eay
where $\gamma$ and $\beta$ are the Lorentz dilation factor and the
velocity of the decaying kaon respectively,
\beq
\gamma=\frac{E_K}{m_K}\,,\quad\quad\quad
\beta=\sqrt{1-\gamma^{-2}}=\frac{|\vec p_K|}{E_K}\;,
\eeq
and $E_0$, $|\vec p_0|$ are defined as
\beq
E_0=\frac{m_K}{2}\left(1-\frac{m_K^2}{m_\pi^2}\right)
-\frac{m_\mu m_K}{m_\pi} \,,\quad\quad |\vec p_0|=\sqrt{E_0^2-m_\mu^2}\;.
\eeq

\begin{figure}[htbp]
\begin{center}
\vspace{1cm}
\epsfig{file=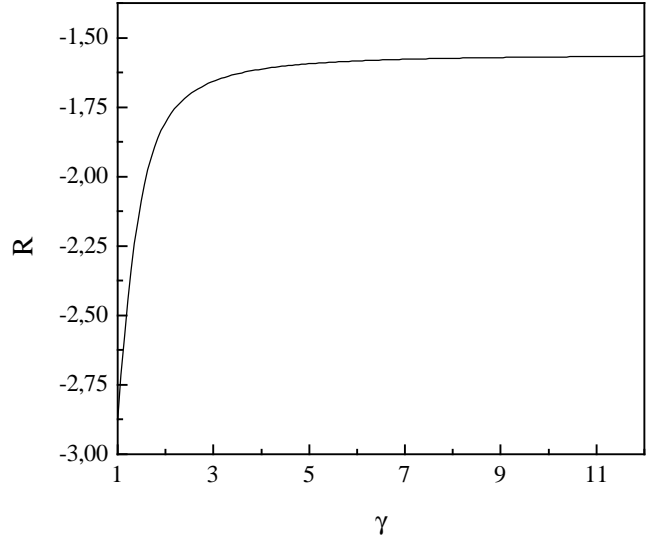}
\end{center}
\caption{Asymmetry parameter $R$, as function of the Lorentz factor
$\gamma$.}
\end{figure}

\hfill

\vspace{0.5cm}

\begin{figure}[htbp] 
\begin{center} 
\epsfig{file=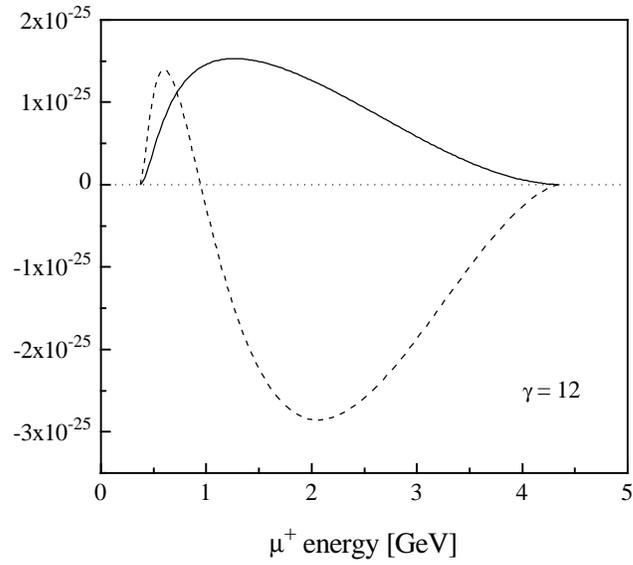}
\end{center} 
\caption{Differential decay rates for the process $\kpmm$, as functions of
the energy of the final $\mu^+$, for a Lorentz factor $\gamma=12$.
The solid line stands for the total width, while the dashed one
corresponds ---up to a global sign--- to the difference between
the rates for opposite longitudinal $\mu^+$ polarizations, scaled
by a factor $({\rm Re}\xi)^{-1}$.}
\label{fig2} 
\end{figure} 
 
\end{document}